\documentclass[showpacs,aps,amssymb,floatfix,prd,amsmath,preprintnumbers]{revtex4}
\usepackage{epstopdf}
\usepackage{capt-of}
\usepackage{graphicx}  
\usepackage{dcolumn}   
\usepackage{bm}
\usepackage{amsmath}
\usepackage[font=scriptsize]{caption}
\usepackage[colorlinks]{hyperref}
\begin{document}
\title{\bf  Axially magnetized Dark Energy cosmological model}

\author{B. Mishra\footnote{Department of Mathematics,Birla Institute of Technology and Science-Pilani, Hyderabad Campus, Hyderabad-500078, India, E-mail:bivudutta@yahoo.com}, Pratik P Ray\footnote{Department of Mathematics, Birla Institute of Technology and Science-Pilani, Hyderabad Campus, Hyderabad-500078, India, E-mail: pratik.chika9876@gmail.com}, S.K. Tripathy \footnote{Department of Physics, Indira Gandhi Institute of Technology, Sarang, Dhenkanal, Odisha-759146, India, E-mail:tripathy\_ sunil@rediffmail.com}, Kazuharu Bamba \footnote{Division of Human Support System, Faculty of Symbiotic Systems Science, Fukushima University, Fukushima 960-1296, Japan, Email:  bamba@sss.fukushima-u.ac.jp}} 

\affiliation{ }

\begin{abstract}
\textbf{Abstract}
We investigate the behaviour of the skewness parameters for an anisotropic universe in the framework of General Relativity. Non interacting dark energy is considered in presence of electromagnetic field. A time varying deceleration parameter simulated by a hybrid scale factor is considered. The dynamics of the universe is investigated in presence and absence of magnetic field. The equation of state parameter of dark energy evolves within the range predicted by the observations. Magnetic field is observed to have a substantial effect on the cosmic dynamics and the skewness parameters. The models discussed here end in a big rip and become isotropic at finite time.

\end{abstract}

\maketitle

\section{Introduction}
Contrary to Newtonian gravity, the recent development of expanding universe is an interesting phenomena in modern cosmology \cite{Hawley98}. Possibly this expansion is due to the Dark Energy(DE), which is instrumental in driving the galaxies apart \cite{Ryden03}. In order to explain acceleration, cosmologists developed the idea of DE, which can be conceived as a hypothetical fluid with negative pressure \cite{Knop03}. Observational evidences \cite{Perlm98, Riess98,Riess04, Ade14, Ade16, Aghanim18} indicate that DE has two-third contribution to the mass energy budget of the universe and has spread across the space homogeneously. The noting part of this discovery is that, DE does not interact with light and remains ineffective on the objects of small scales. Since the DE has a negative pressure, strong energy condition is not satisfied. For this reason, the evolution of the universe gets a sudden boost and there occurs a transition from early deceleration to a late time acceleration. It can also be noted that, cosmologists have made several attempts to measure the deceleration of the expansion with a strong belief that the expansion of the universe due to gravity may slow down. However, the observational results from supernovae observations, High-z supernovae search team totally rejected this idea and almost proved that the expansion of the universe is accelerating \cite{Riess98, Perlm99}. Subsequently strong evidences were gathered in support of the accelerating expansion with greater accuracy \cite{Riess00, Astier06, Amanullah10, Weinberg13}.  Later, the fact is supported by other observations from BAO \cite{Bond97, Wang06}, SDSS \cite{Seljak05, Adelman06}, CMB \cite{Jaffe06, Spergel07}. However, definitive observation of dark energy has remains to be elusive.\\

Post supernovae observations, plethora of cosmological models based on accelerating expansion of the universe have been proposed either by modifying the matter part of Einstein's Field Equations or by modifying the gravity theory. Bamba et al. \cite{Bamba12} have explained the dark energy and dark matter by a two periodic generalizations of the Chaplygin gas type models. Brevik et al. \cite{Brevik15} constructed cosmological models by establishing an interaction between the DE and dark matter with a homogeneous equation of state parameter. In contract to constant deceleration parameter, Mishra and Tripathy \cite{Mishra15b} considered time varying deceleration parameter to construct DE cosmological model in an anisotropic metric. Saadeh et al. \cite{Saadeh16} have presented a new framework to search for any departures from isotropic background in Bianchi type $VII_h$ cosmologies. Fayaz and Hossienkhani \cite{Fayaz17} have considered Bianchi I space time and made a comparison between generalized Chaplygin gas model, interacting holographic and new agegraphic dark energy. Mamon et al. \cite{Mamon17} have considered dimensionless dark energy function to reconstruct the DE cosmological model. Ebrahimi et al. \cite{Ebrahimi17} have established a non-linear interaction of dark energy and dark matter to construct ghost DE cosmological model. Several cosmological models have been constructed  to address the cosmic acceleration problem using different space times and different matter fields \cite{Bamba11, Bamba12a, Mishra17, Mishra18}. Chiral cosmological models as generalization of multicomponent scalar field have also been proposed to understand the mechanism of late time cosmic speed up phenomenon\cite{abbya2012, abbya2015, Chervon}. Recent works on dark energy and the extended gravity can  be found in \cite{Capozziello:2010zz,Nojiri:2010wj,Bamba:2012cp,Bamba:2015uma, Cai:2015emx,Nojiri:2017ncd}.

The universe is observed to be isotropic. But there continues a debate on whether Bianchi type models are favoured at some cosmic phases or not? On the theoretical front, spatially homogeneous and isotropic FRW are usually considered for modelling the present universe. However, Bianchi type models with flat spatial sections and anisotropic directional expansion rates are considered more general compared to FRW models. In the present work, we have employed Bianchi-V space time to construct the model. The motivation is to develop a more general model which can allow us to consider anisotropy of any range (comparable to the range predicted in observations). In most of the cosmological models, we have observed that the deceleration parameter assumed to be constant and the matter field has been taken either as perfect fluid or ordinary matter. However with this background, it is not sufficient to observe the dynamical behaviour with anisotropy for an accelerating universe. In this context, to get into the deeper inside of the problem, we consider the combination of two fluids such as the DE fluid and the fluid of usual matter. Akarsu and Kilinc \cite{Akarsu10, Akarsu11}, assumed constant deceleration parameter to investigate the DE problem with the anisotropic metric Bianchi I and Bianchi III. Yadav et al. \cite{Yadav11} have supported their DE cosmological model with variable deceleration parameter and anisotropic Bianchi V space time. Theoretical models of interacting and non interacting DE have been discussed widely in the literature \cite{Mishra17, Mishra18, Mishra17a, Mishra18a}. Several cosmological models have been constructed on cosmic magnetic field as the matter field \cite{lebla, tsaq, skt2009, skt15}. The role of cosmic magnetic fields has been reviewed in Refs. \cite{Kronberg:1993vk, Grasso:2000wj, Carilli:2001hj, Widrow:2002ud, Giovannini:2003yn, Giovannini:2004rj, Giovannini:2006kg, Kandus:2010nw, Yamazaki:2012pg, Durrer:2013pga, Maleknejad:2012fw, Bamba:2003av, Bamba:2004cu,Bamba:2006ga,Bamba:2008ja,Bamba:2008xa,Bamba:2018cup}. 

The paper has been arranged as follows: in section II, we have developed the mathematical formalism for Bianchi type V space-time in a two-fluid situation and derived its physical parameters. In section III, the solution of the model is obtained with a hybrid scale factor. In section IV, a comparison is made for EoS and skewness parameters with respect to different representative values and the evolutionary aspect of the skewness parameters has been discussed. The existence of big bang and big rip singularity has also been analysed in this section. Concluding remarks have been presented in section V. 

\section{Mathematical Formalism with Physical Parameters}

In the context of the recent observational results made us to believe that, the universe at least at the late phase of cosmic evolution undergoes accelerating expansion. Further, it is mentioned that this result is due to the presence of an exotic DE energy form. Hence, in order to study the accelerating expansion further, we have considered here the cosmic fluid composed of DE and electromagnetic field. The motivation behind such an assumption stems from the fact that, DE has a lion share of $68.3\%$ in the mass-energy budget compared to a negligible share of around $4.9\%$ for the baryonic matter. The presence of electromagnetic field may also contribute to some extent to the cosmic speed up phenomena. Also, presence of magnetic field leads to an asymmetric cosmic expansion. Now, the energy momentum tensor(EMT)for the cosmic fluid can be expressed as 

\begin{equation}
T_{\mu \nu}=E_{\mu \nu}+T_{\mu \nu}^{D}, \label{eq:1}
\end{equation}
where $E_{\mu \nu}$ is the energy momentum tensor for electromagnetic field whereas $T_{\mu \nu}^{D}$ is for the DE. For this investigation, we have considered an anisotropic space time in the form of Bianchi type V (BV) metric as 
  \begin{equation} \label{eq:2}
 ds^{2}= dt^{2}-a(t)^{2} dx^{2}-e^{2\alpha x}[b(t)^{2} dy^{2}+c(t)^{2} dz^{2}]
\end{equation}
In BV space time \eqref{eq:2}, it is seen that the expansion along different directions are different for the change in the cosmic time. The time dependent functions $a(t), b(t)$ and $ c(t)$ change differently with cosmic time and are dubbed as the directional scale factors respectively for $x$, $y$ and $z$ direction. $\alpha$ is non-zero positive constant. It is obvious that, when the three time dependent functions are equal and the constant $\alpha=0,$ the line element \eqref{eq:2} reduces to flat FRW metric. In this sense, the BV metric is a general model than FRW model where some scope for anisotropic expansion can be assumed. Another geometrical significance of \eqref{eq:2} is that  the expansion along the longitudinal direction $x$ is different than that of the two traverse directions. In order to study an anisotropic cosmological model, we should have an anisotropic distribution of the DE fluid. So, the energy momentum tensor for DE can be defined as

\begin{align} \label{eq:3}
T^{D}_{\mu \nu}  &=  diag[\rho, -p_{x},-p_{y},-p_{z}]\\ \nonumber
			        &= diag[\rho, -\rho\omega_{x},-\rho\omega_{y},-\rho\omega_{z}]\\ \nonumber
			        &= diag[\rho, -\rho(\omega+\delta), -\rho(\omega+\gamma),-\rho(\omega+\eta)]
\end{align}
where $p$ is the pressure, $\rho$ is the density of DE and the DE equation of state parameter (EoS), $\omega=\frac{p}{\rho}$. Since the pressure anisotropy is one of the source of anisotropy, we have expressed that in \eqref{eq:3} through the skewness parameters $\delta$, $\gamma$ and $\eta$. These parameters are the deviations from the EoS parameter along the respective coordinate axis. As we have added the electromagnetic field with DE, we can define its energy momentum tensor in the form 

\begin{equation}\label{eq:4} 
E_{\mu \nu} = \frac{1}{4 \pi} \left[ g^{sp}f_{\mu s}f_{\nu p}-\frac{1}{4} g_{\mu \nu}f_{sp}f^{sp}   \right],
\end{equation}
where $g_{\mu \nu}$ is the gravitational metric potential and $f_{\mu \nu}$ is the electromagnetic field tensor. In order to avoid the interference of electric field, we have considered an infinite electrical conductivity to construct the cosmological model. This results in the expression $f_{14}=f_{24}=f_{34}=0$. Again, quantizing the axis of the magnetic field along $x$-direction as the axis of symmetry, we obtain  $f_{12}=f_{13}=0,$ $f_{23}\neq 0$. Thus, the only non-vanishing component of electromagnetic field tensor is $f_{23}$.\\

With the help of Maxwell's equation, the non-vanishing component can be represented as, $f_{23}=-f_{32}= k $, where  $k$ is assumed to be a constant and it comes from the axial magnetic field distribution. For the anisotropic BV model, the components of EMT for electromagnetic field can be obtained as

\begin{align}\label{eq:5}
E_{11} &= \mathcal{M} a^{2},\nonumber\\
E_{22} &=- \mathcal{M} b^{2} e^{2 \alpha x},\nonumber\\
E_{33} &=- \mathcal{M} c^{2} e^{2 \alpha x},\nonumber\\
E_{44} &=- \mathcal{M},
\end{align}
where, $\mathcal{M}=\frac{k^{2}}{8 \pi b^{2}c^{2}e^{4 \alpha x}}$. When $k$ is non-zero, the presence of magnetic field has been established along the $x$-direction. If $k$ vanishes, the model will reduce to the one with DE components only.

Assuming GR is well defined at cosmic scales, we wish to investigate the axially magnetised universe in the framework of Einstein's field equations (EFE),
\begin{equation} \label{eq:6}
R_{ij}-\frac{1}{2}Rg_{ij}=\kappa T_{ij}.
\end{equation}
In \eqref{eq:6}, $R_{ij}$ is Ricci tensor and the Ricci scalar, $R=g_{ij}R^{ij}$, $\kappa = - \frac{8 \pi G}{c^{4}}$ . For physical convenience, we adopt  natural units where, $G =\frac{1}{8\pi}$ and $c=1.$ For the assumed cosmological model, the Einstein field equations become

\begin{align}
& \frac{\ddot{b}}{b}+\frac{\ddot{c}}{c}+\frac{\dot{b}%
\dot{c}}{bc}-\frac{\alpha^{2}}{b^{2}}=-(\omega+\delta)\rho- \mathcal{M}  \label{eq:7} \\
& \frac{\ddot{a}}{a}+\frac{\ddot{c}}{c}+\frac{\dot{a}%
\dot{c}}{ac}-\frac{\alpha^{2}}{a^{2}}=-(\omega+\gamma)\rho+\mathcal{M}  \label{eq:8} \\
& \frac{\ddot{a}}{a}+\frac{\ddot{b}}{b}+\frac{\dot{a}%
\dot{b}}{ab}-\frac{\alpha^{2}}{a^{2}}=-(\omega+\eta)\rho+ \mathcal{M}  \label{eq:9} \\
& \frac{\dot{a}\dot{b}}{ab}+\frac{\dot{b}\dot{c}}{%
bc}+\frac{\dot{c}\dot{a}}{ca}-\frac{3\alpha^{2}}{a^{2}%
}= \rho- \mathcal{M}  \label{eq:10} \\
& 2\dfrac{\dot{a}}{a}-\dfrac{\dot{b}}{b}-\dfrac{\dot{c}}{%
c} =0.  \label{eq:11}
\end{align}

The single and double over dot on the metric potentials respectively denote first and second derivatives with respect to the cosmic time. Field equation \eqref{eq:11} yields $a^2=bc$ after absorbing the integration constant into the metric potential $b$ or $c$.

In order to analyse the model, we define the spatial volume with respect to the metric potentials as $V=R^3=abc$ whereas the scalar expansion can be expressed as $\theta=3H$. Moreover, the scale factor can be denoted as $\mathcal{R}(t)$. Now, we can also establish relationship between the direction Hubble rates ($H_{x},H_{y},H_{z}$), mean Hubble rate ($H$) and the scale factor ($\mathcal{R}$)as $H= \frac{\dot{\mathcal{R}}}{\mathcal{R}}= \frac{1}{3}(H_{x}+H_{y}+H_{z})$.
The shear scalar is given by,
$\sigma^2=\frac{1}{2}(H_x^2+H_y^2+H_z^2-\frac{\theta^2}{3})$; $ q=-\frac{\mathcal{R}\ddot{\mathcal{R}}}{\dot{\mathcal{R}}^2} $ is the deceleration parameter. \\

Incorporating the energy conservation, $T^{\mu \nu}_{; \nu}=0,$ for the magnetized DE fluid, we obtain

\begin{equation} \label{eq:12}
\dot{\mathcal{M}}+2 \mathcal{M} (H_{y}+H_{z})+\dot{\rho}+3\rho[(\omega+1)H+(\delta H_x+\gamma H_y+\eta H_z)/3]=0.
\end{equation}
For a mixture of non interacting DE fluid and electromagnetic fluid, separate energy conservation equations can be obtained from \eqref{eq:12} as

\begin{equation} \label{eq:13}
\dot{\mathcal{M}}+2 \mathcal{M} (H_{y}+H_{z})=0
\end{equation}
and
\begin{equation} \label{eq:14}
\dot{\rho}+3\rho[(\omega+1)H+(\delta H_x+\gamma H_y+\eta H_z)/3]=0
\end{equation}

Integrating eq. \eqref{eq:13}, we obtain
\begin{equation} \label{eq:15}
\mathcal{M}= \frac{k_{1}}{b^{2}c^{2}},
\end{equation}
where $k_{1}$ is an integration constant and we can assign $k_{1}=\frac{k^{2}}{8 \pi e^{4 \alpha x}}$ to retain the value of $\mathcal{M}$ as obtained in eq.\eqref{eq:5} without affecting its generality. 

The first two terms of eq.\eqref{eq:14} are the deviation free parts of the EoS parameter whereas the third term is the respective deviations of the EoS parameter. Now, we can obtain

\begin{equation} \label{eq:16}
\frac{\dot{\rho}}{\rho}= -3(\omega+1)\frac{\dot{\mathcal{R}}}{\mathcal{R}},
\end{equation}
and
\begin{equation} \label{eq:17}
\delta=-\frac{\gamma H_y+\eta H_z}{H_x}.
\end{equation}
It can be inferred from eq. \eqref{eq:16} that the behaviour of dark energy density $(\rho)$ is decided by the time evolution of the deviation free DE EoS parameter. Eq. \eqref{eq:17} deals with the deviation of DE pressure along different coordinate axes.\\

To investigate the dynamical and anisotropic behaviour along different directions, we adopt here the mathematical formalism developed in \cite{Mishra15b}. Now, with an axially placed magnetic field and DE, we reveal the dynamical properties of the cosmological model. To do that we have taken an anisotropic relationship between the metric potentials as $b=c^m$ \cite{Mishra15b,Mishra17a}. Here $m$ should not be  unity in order to avoid the isoptropic behaviour of the model. The anisotropic parameter $\cal(A)$ can be obtained as, ${\cal A}=\frac{1}{3}\sum \left(\frac{\vartriangle H_i}{H}\right)^2= \frac{2}{3} \left(\frac{m-1}{m+1} \right)^{2}$. With this, the metric potentials with respect to scale factors are, $a=\mathcal{R}, b=\mathcal{R}^{\frac{2m}{m+1}}, c=\mathcal{R}^{\frac{2}{m+1}}$ respectively. The directional Hubble parameters changed its expressions as, $H_x=H$, $H_y=\left(\frac{2m}{m+1}\right)H$ and $H_z=\left(\frac{2}{m+1}\right)H$. 

Now, with algebraic manipulation of equations \eqref{eq:7} ,\eqref{eq:8}, \eqref{eq:9} and \eqref{eq:17}, the skewness parameters take the form,

\begin{eqnarray}
\delta &=& -\frac{2}{3\rho}\left[\epsilon_1(m)F(\mathcal{R})+\mathcal{M}\right],\label{eq:18}\\
\gamma &=&  \frac{1}{3\rho}\left[\epsilon_2(m)F(\mathcal{R})+\mathcal{M}\right],\label{eq:19}\\
\eta   &=& -\frac{1}{3\rho}\left[\epsilon_3(m)F(\mathcal{R})-\mathcal{M}\right]\label{eq:20}.
\end{eqnarray}
Here $F(\mathcal{R})=\biggl( \dfrac{\ddot{\mathcal{R}}}{\mathcal{R}}+ \dfrac{2 \dot{\mathcal{R}}^{2}}{\mathcal{R}^{2}} \biggr)$ and the $m$ dependent functionals are $\epsilon_1(m)=\left(\frac{m-1}{m+1}\right)^2$, 
$\epsilon_2(m)=\left(\frac{m+5}{m+1}\right)\left(\frac{m-1}{1+m}\right)$ and 
$\epsilon_3(m)=\left(\frac{5m+1}{m+1}\right)\left(\frac{m-1}{1+m}\right)$.  Further, the skewness parameters are scale factor dependent; specifically on the functional $F(\mathcal{R})$, the magnetic contribution $\mathcal{M}$  and the DE energy density $\rho.$ For an isotropic model, $\epsilon_1(m),\epsilon_2(m), \epsilon_3(m) $ vanish and consequently the skewness parameters become proportional to the ratio $\frac{\mathcal{M}}{\rho}$ i.e  $\delta=-\frac{4}{3}\frac{\mathcal{M}}{\rho}$, $\gamma=\frac{2}{3}\frac{\mathcal{M}}{\rho}$ and $\eta=\frac{2}{3}\frac{\mathcal{M}}{\rho}$. It is observed that, the presence of axial magnetic field contributes to DE pressure anisotropy along different coordinates. The above relations reduce to the respective expressions obtained in \cite{Mishra15b} in the absence of any magnetic field.

The dark energy density $\rho$ and the EoS parameter $\omega$ are obtained as,

\begin{eqnarray} 
\rho &=& 2 \phi_{1} (m) \dfrac{\dot{\mathcal{R}}^2}{\mathcal{R}^2}- 3 \dfrac{\alpha ^2}{\mathcal{R}^2}+ 2\mathcal{M},\label{eq:21}\\
\omega\rho &=& \left(\frac{\mathcal{A}}{2}-1\right)\left[\dfrac{2\ddot{\mathcal{R}}}{ \mathcal{R}}+ \dfrac{\dot{\mathcal{R}}^2}{\mathcal{R}^2} \right] + \frac{2}{3}\mathcal{M}+ \dfrac{\alpha ^{2}}{\mathcal{R}^2}, \label{eq:22}
\end{eqnarray}

where, $\phi_{1} (m)= \dfrac{ (m^{2} + 4 m + 1)}{(m+1)^2}$. 

From eq. \eqref{eq:21}-\eqref{eq:22}, the importance of the exponent $m$ is clearly visible in the sense that evolutionary behaviour of the EoS parameter can be assessed through this exponent $m$, the average scale factor $\mathcal{R}$, the magnetic contribution $\mathcal{M}$ and the rest energy density $\rho$. Magnetic field certainly affects the evolutionary behaviour of the DE EoS parameter. In the absence of the axial magnetic field, the above equations reduce to the respective expressions of \cite{Mishra15b}.

\section{Hybrid scale factor and Dynamical aspects of the model}
We are indeed focussing on the variable deceleration parameter to understand the anisotropic behaviour on the cosmological models. This is because there is a strong believe that universe evolved from a decelerated phase of expansion to an accelerated phase. This can be better studied if the deceleration parameter is time dependent. Keeping this in mind, we have chosen a hybrid scale factor in the form (HSF) \cite{Mishra15b, Mishra17a,Mishra18d} 
\begin{equation}
\mathcal{R} = e^{h_{1}t}t^{h_{2}}\label{eq:23}
\end{equation}
where $h_{1}$ and $h_{2}$ are two adjustable parameters. With this type of scale factor,  we can define the corresponding Hubble parameter as  $H = \left( h_{1}+\frac{h_{2}}{t}\right)$. One should note that the deceleration parameter $q$  is an important quantity in describing cosmic dynamics of universe.  Positive value of it indicates decelerating universe where as negative values favour a universe with accelerated expansion. With the advent of a lot of observations from high redshift type Ia supernova combined with BAO and CMB, models  predicting a transition from early decelerating universe to late time accelerating universe gained much importance in recent times. At the present epoch, the negative value of $q$ such as $q = -0.81 \pm 0.14$ is more favourable according to recent observational data \cite{Ade14, Ade16}. Supporting to this fact, the deceleration parameter for the present model, incorporating hybrid scale factor, turns out to be $ q = - 1 + \frac{h_{2}}{(h_{1}t+h_{2})^{2}}$. At early phase of evolution $(t\rightarrow 0)$ and $q$ tends to be $-1+\dfrac{1}{h_2}$ where as $q \simeq -1 $ at late phase $(t \rightarrow \infty).$ The parameter $h_2$ can be constrained in the range $0 < h_{2} < 1$ from simple arguments; however with a detail analysis of the Hubble parameter with respect to redshift data, the range for $h_1$ can be constrained \cite{Mishra15b}. Hence, here we have considered the range for $h_{2}$ as in  \cite{Mishra15b} and $h_{1}$ as an open parameter. The directional scale factors or metric potentials for an HSF can be expressed as  $a=\mathcal{R}=e^{h_{1}t}t^{h_{2}}, b=(e^{h_{1}t}t^{h_{2}})^{\frac{2m}{m+1}} $ and $c=(e^{h_{1}t}t^{h_{2}})^{\frac{2}{m+1}}$.

For the present model with an HSF, the pressure anisotropies along different spatial directions are obtained as
\begin{eqnarray}
\delta &=& -\frac{2}{3\rho}\left[\epsilon_1(m)F(t)+\mathcal{M}\right],\label{eq:24}\\
\gamma &=&  \frac{1}{3\rho}\left[\epsilon_2(m)F(t)+\mathcal{M}\right],\label{eq:25}\\
\eta   &=& -\frac{1}{3\rho}\left[\epsilon_3(m)F(t)-\mathcal{M}\right]\label{eq:26}.
\end{eqnarray}
where
\begin{equation}
F(t)= \frac{3h_1^2t^2+6h_1h_2t+3h_2^2-h_2}{t^2}.\label{eq:27}
\end{equation}

The DE density and EoS parameter for the present model with an HSF can also be obtained in a straightforward manner using eqs. \eqref{eq:21}-\eqref{eq:23}. In the present model, we wish to put a positive constraint on the dark energy density which forces us to chose low values of the parameter $\alpha$ so that the dark energy density $\rho$ remains positive through out the cosmic evolution. This constraint enables the model to satisfy weak energy condition (WEC) and  null energy condition (NEC). With all such constraints on the model parameters, the energy density $\rho$ decreases with an increase in cosmic time and reaches to a constant positive value at late epoch. 

\section{Results and Physical Behaviours of the Model}
The model as developed in this work depends on the model parameters. Primarily five parameters  are responsible in deciding the dynamical behaviour of the model. Two parameters $h_1, h_2$ from the hybrid scale factor, $\alpha$ from the anisotropic BV metric, anisotropic parameter $m$ and $k$ from the strength of the axial magnetic field. As we have mentioned earlier, the parameter $h_1$ is considered to be within the range $0<h_1<1$ and $h_2$ is taken as a free parameter. In order to investigate various dynamical aspects of the model particularly the evolution of the equation of state parameter and the directional pressure anisotropies, we have considered $h_1=0.1$ and $h_2=0.4$. These choices are motivated from the cosmic transit phenomena where we expect that the deceleration parameter smoothly decreases from a positive value at an early epoch to negative values at late times and can predict a transit redshift of the order of unity ($z_t \simeq 1$). We have considered low values of $\alpha$ in the range $0<\alpha<0.5$ so that the dark energy density remains in the positive domain. The anisotropic parameter $m$ has been constrained in an earlier work to be $m=1.0001633$ which corresponds to an acceptable range of anisotropy \cite{Mishra15b}. The magnetic field contribution is suitably handled in the work to assess the effects of magnetic field on the equation of state parameter and the skewness parameters. The values of these parameters are mentioned (wherever necessary) in the figures.

\subsection{EoS Parameter}
\begin{figure}[h!]
\begin{center}
\includegraphics[width=0.8\textwidth]{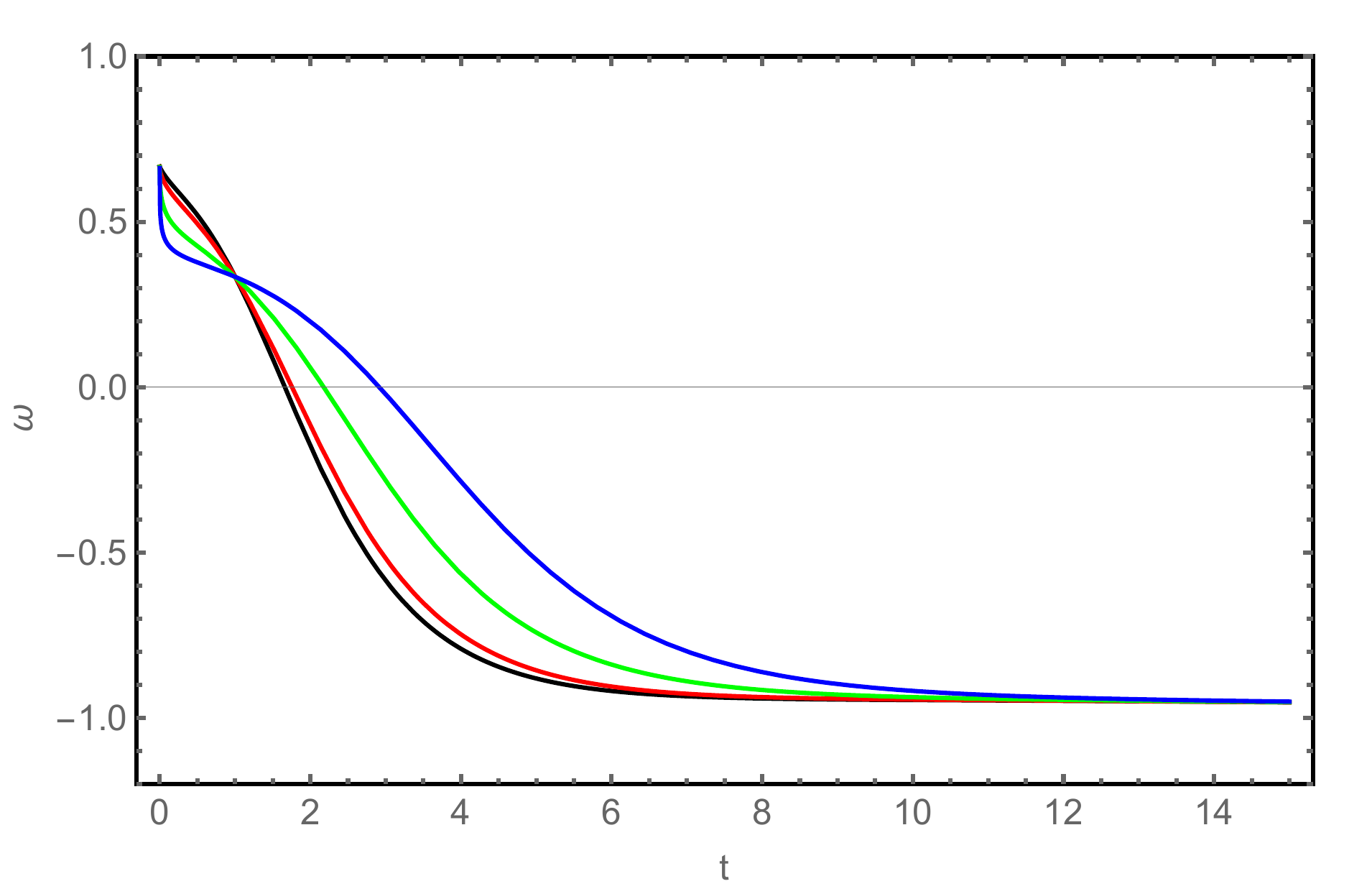}
\caption{EoS parameter for different representative values of the parameter $k$ for $\alpha=0.35$.}
\end{center}
\end{figure}

\begin{figure}
\minipage{0.40\textwidth}
\centering
\includegraphics[width=65mm]{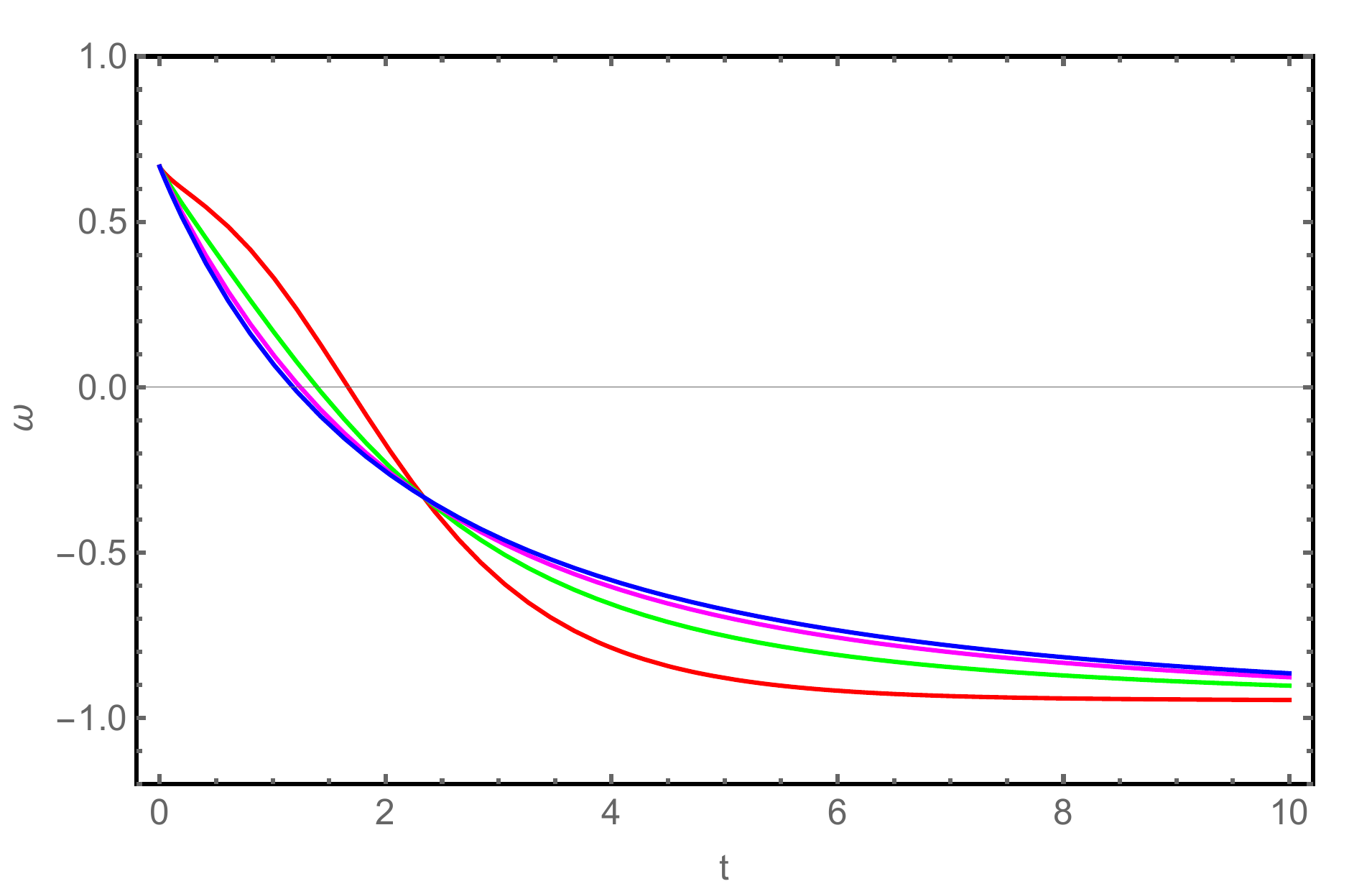}
\caption{EoS parameter for four representative values of the parameter $\alpha$ for $k=0.5$. } 
\endminipage\hfill
\minipage{0.40\textwidth}
\includegraphics[width=65mm]{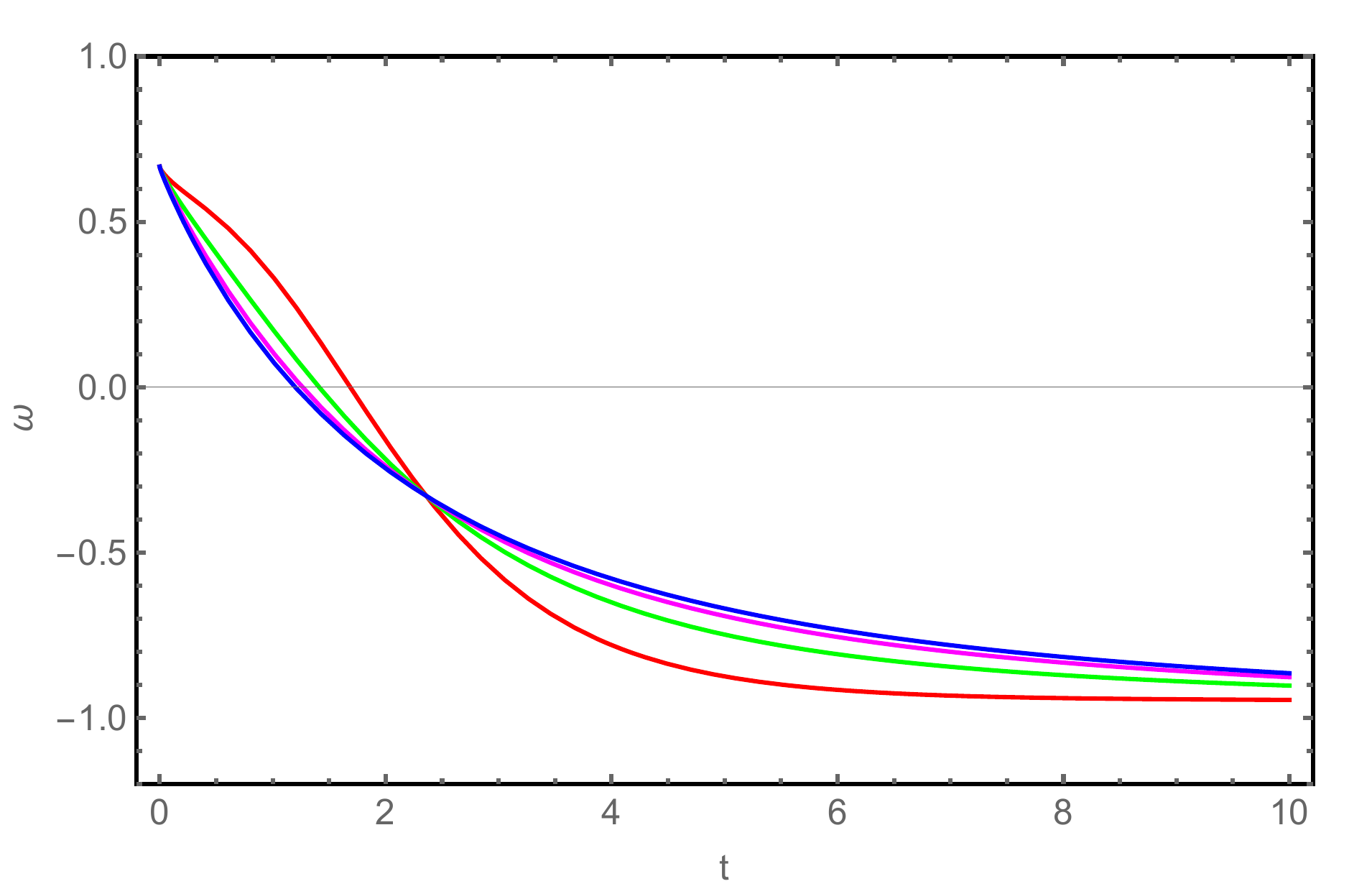}
\caption{EoS parameter for four representative values of the parameter $\alpha$ for $k=1$. }
\endminipage\\
\minipage{0.40\textwidth}
\centering
\includegraphics[width=65mm]{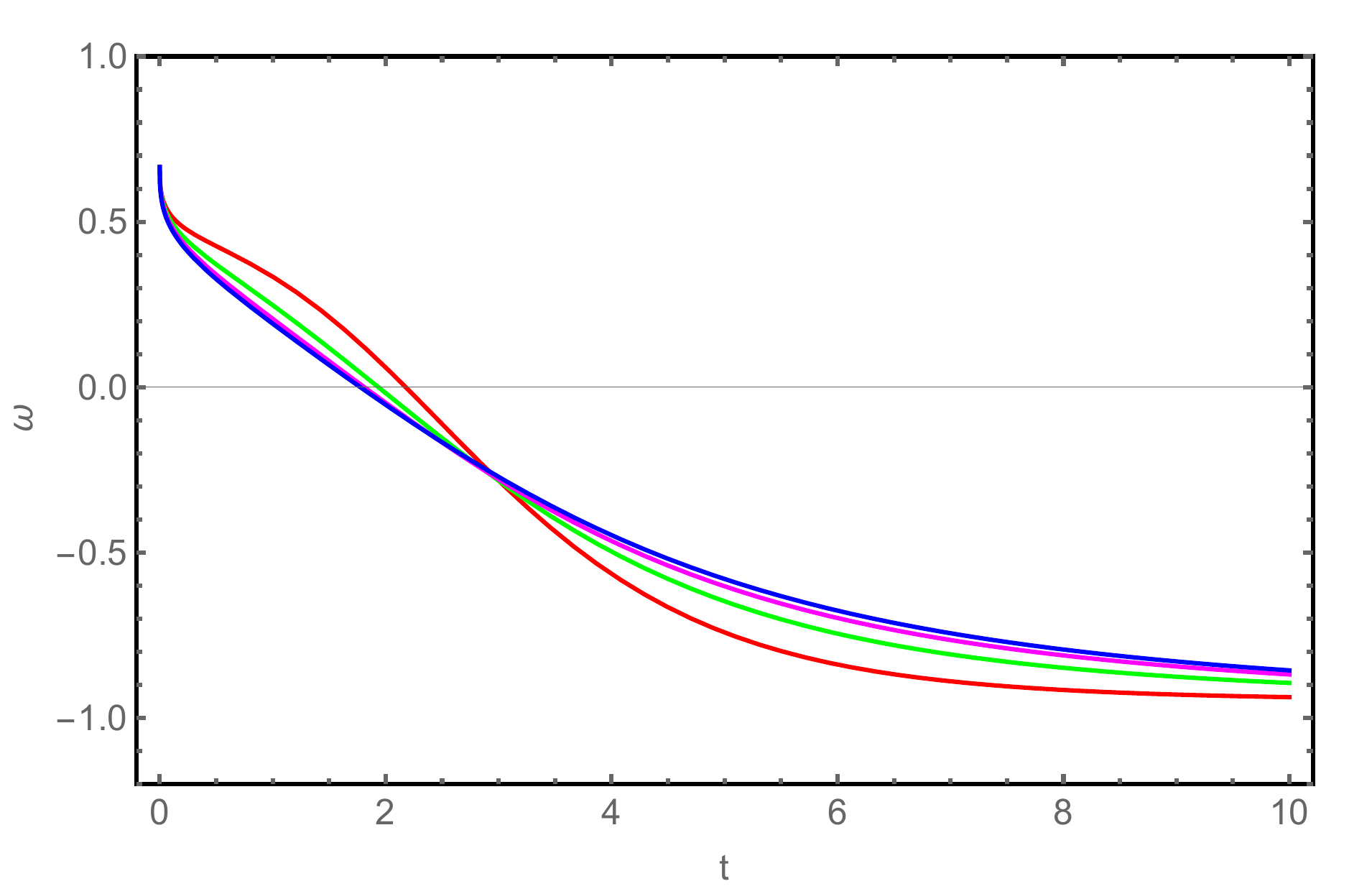}
\caption{EoS parameter for four representative values of the parameter $\alpha$ for $k=5$. } 
\endminipage\hfill
\minipage{0.40\textwidth}
\includegraphics[width=65mm]{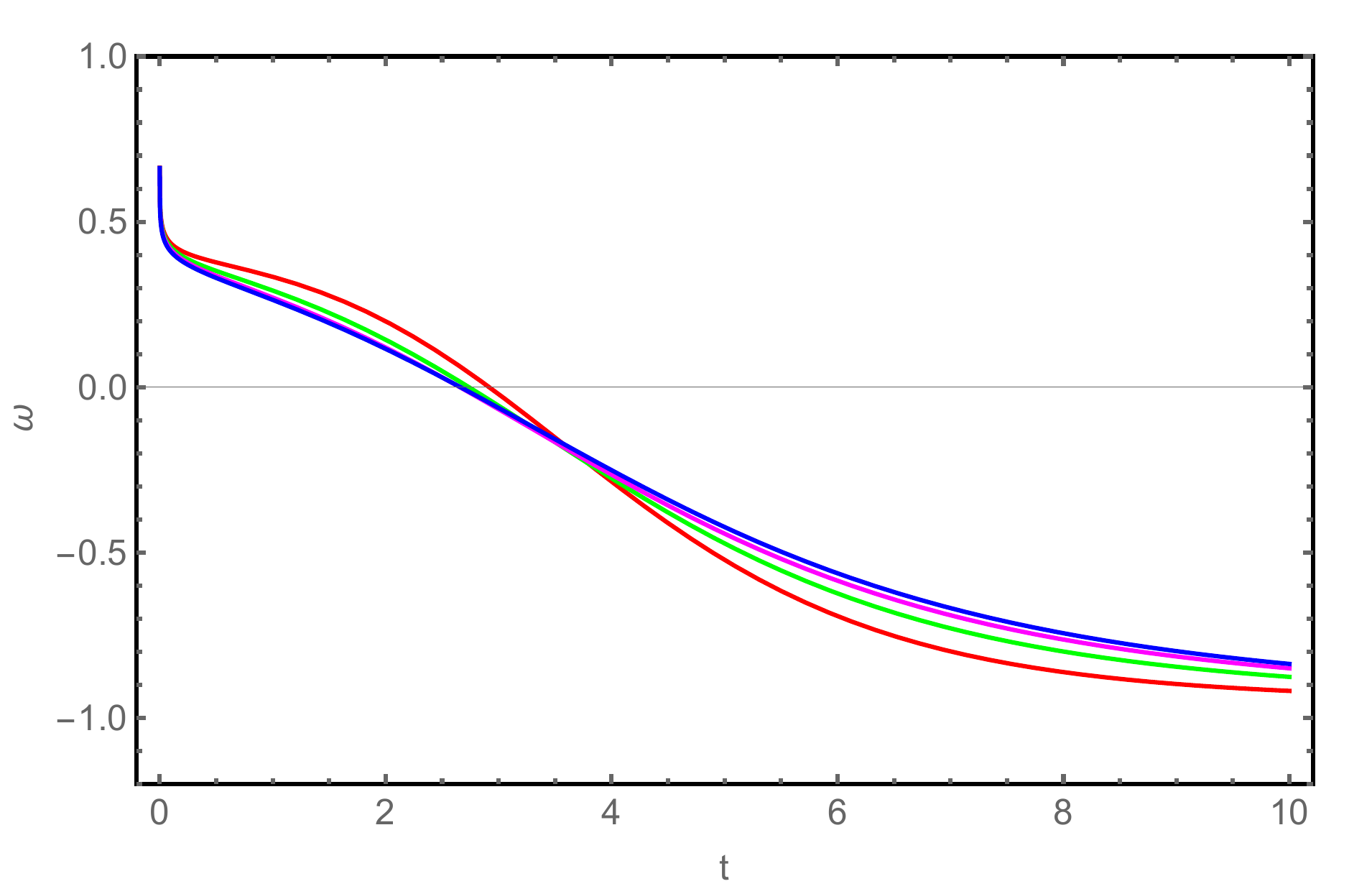}
\caption{EoS parameter for four representative values of the parameter $\alpha$ for $k=9$. }
\endminipage
\end{figure}


In Fig. 1,  the EoS parameter is plotted as a function of cosmic time for different strengths of the magnetic field for a constant value of $\alpha$. It has already been mentioned in the text that, the magnetic field strength is controlled by a parameter $k$. We have considered four different value of $k$ namely $0,2, 5, 9$ to assess its effect on the dynamical evolution of the EoS parameter. In general $\omega$ evolves from a positive domain to a negative domain. At late phase it saturates at around $-1$ to overlap with $\Lambda$CDM model. The equation of state parameter displays peculiar behaviour with the increase in the strength of the axial magnetic field.
At an early epoch, the effect of magnetic field is substantial compared to that at late phase of cosmic evolution. At very late phase, the magnetic field has minimal effect on $\omega$ so that the curves of $\omega$ coincide and saturates at around $\omega=-1$. At an early epoch, the equation of state parameter decreases with the increase in the magnetic field strength. But this behaviour is reversed in due course of cosmic evolution and $\omega$ increases with increase in magnetic field.

In order to assess the effect of the parameter $\alpha$ on the dynamical evolution of the EoS parameter, we have plotted $\omega$ for four representative values of $\alpha$ for given axial magnetic field strength in Fig. 2. The red curve in the plot corresponds to $\alpha=0.35$, the green curve is for $\alpha=0.25$, pink curve for $\alpha=0.15$ and the blue curve represents for $\alpha=0.05$. It is obvious from the plot that, $\omega$ increases with the increase in $\alpha$ in very early epoch and after some cosmic phase of evolution it changes the behaviour. At some cosmic time, $\omega$ reverses the trend and decreases with the increase in $\alpha$. Just like the effect of magnetic field, $\alpha$ affects the dynamics in the early phase of evolution. However $\omega$ becomes less sensitive to $\alpha$ values at late phase. The Figures 3-5 are a repetition of Fig.2 with different $k$ values. From the Figs. 2-5, the effect of the magnetic field on the EoS parameter can be clearly assessed. It is certain from the analysis that, magnetic field has a substantial affect in the early phase of cosmic evolution.

\subsection{Skewness parameters}
The dynamical evolution of pressure anisotropies along different orthogonal spatial directions are investigated in terms of the skewness parameters $\delta, \gamma$ and $\eta$ in Figs. 6-11. In Fig.6, the skewness parameters in the absence of magnetic field are shown. In the absence of magnetic field, the skewness parameters depend on the parameter $m$ and the parameters $h_1, h_2$ of HSF. In the present work, we have considered $m=0.0001633$, $h_1=0.1$ and $h_2=0.4$. The parameters are chosen in such a manner that, we obtain no substantial evolution of the skewness parameter along the x-direction. However, the skewness parameters along the y-direction and z-direction evolve with cosmic evolution. In the absence of magnetic field, $\gamma$ remains to be positive where as $\eta$ becomes negative.  With the cosmic evolution, the magnitude of $\gamma$ and $\eta$ increase for some period of time and then decrease after attaining a maximum. In general, the evolutionary behaviour of $\gamma$ seems to be opposite to that of $\eta$.

\begin{figure}[tbph]
\begin{center}
\minipage{0.32\textwidth}
\centering
\includegraphics[width=\textwidth]{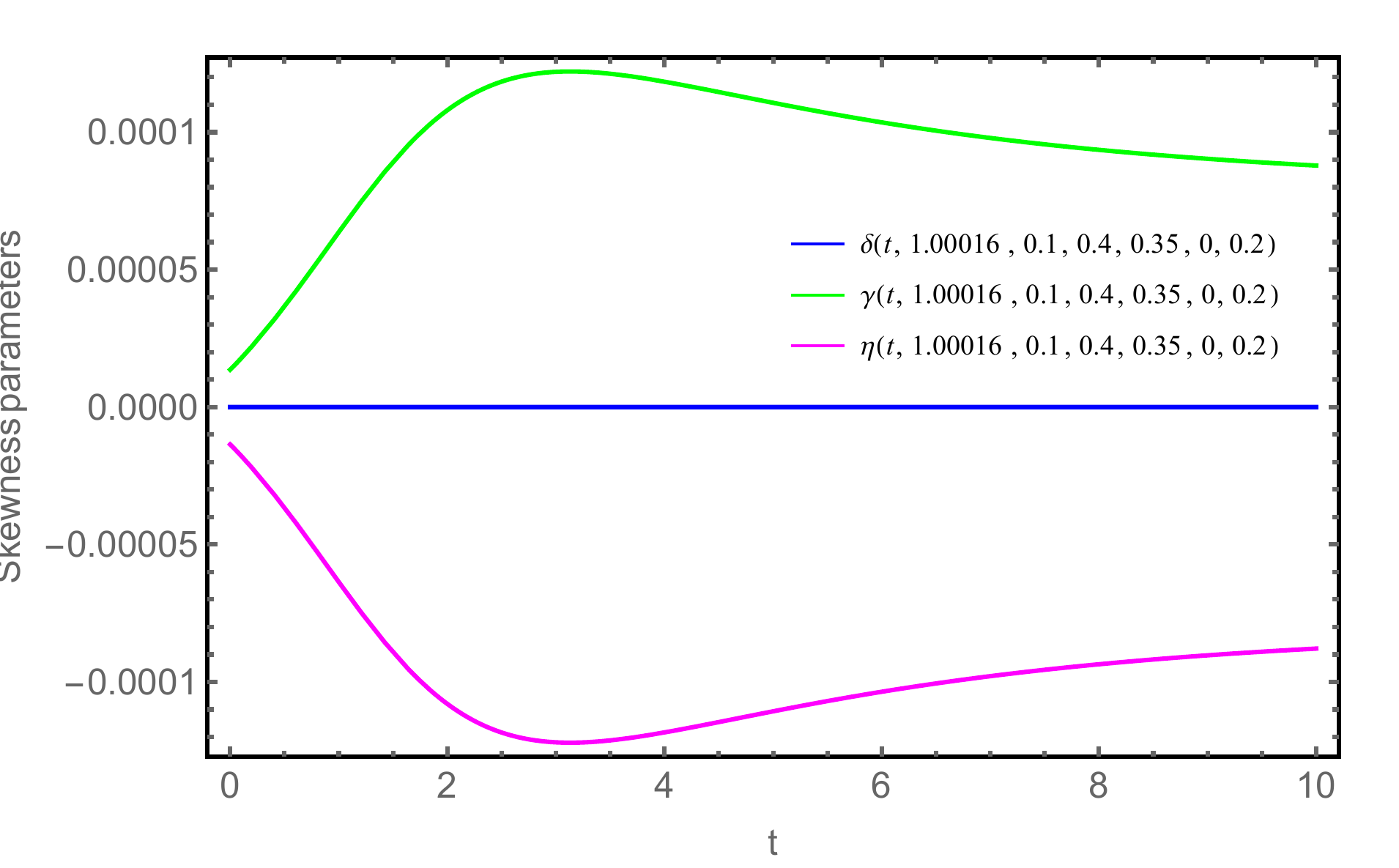}
\caption{Skewness parameters in the absence of magnetic field ($k=0$) with $\alpha=0.35$.}
\endminipage
\minipage{0.32\textwidth}
\centering
\includegraphics[width=\textwidth]{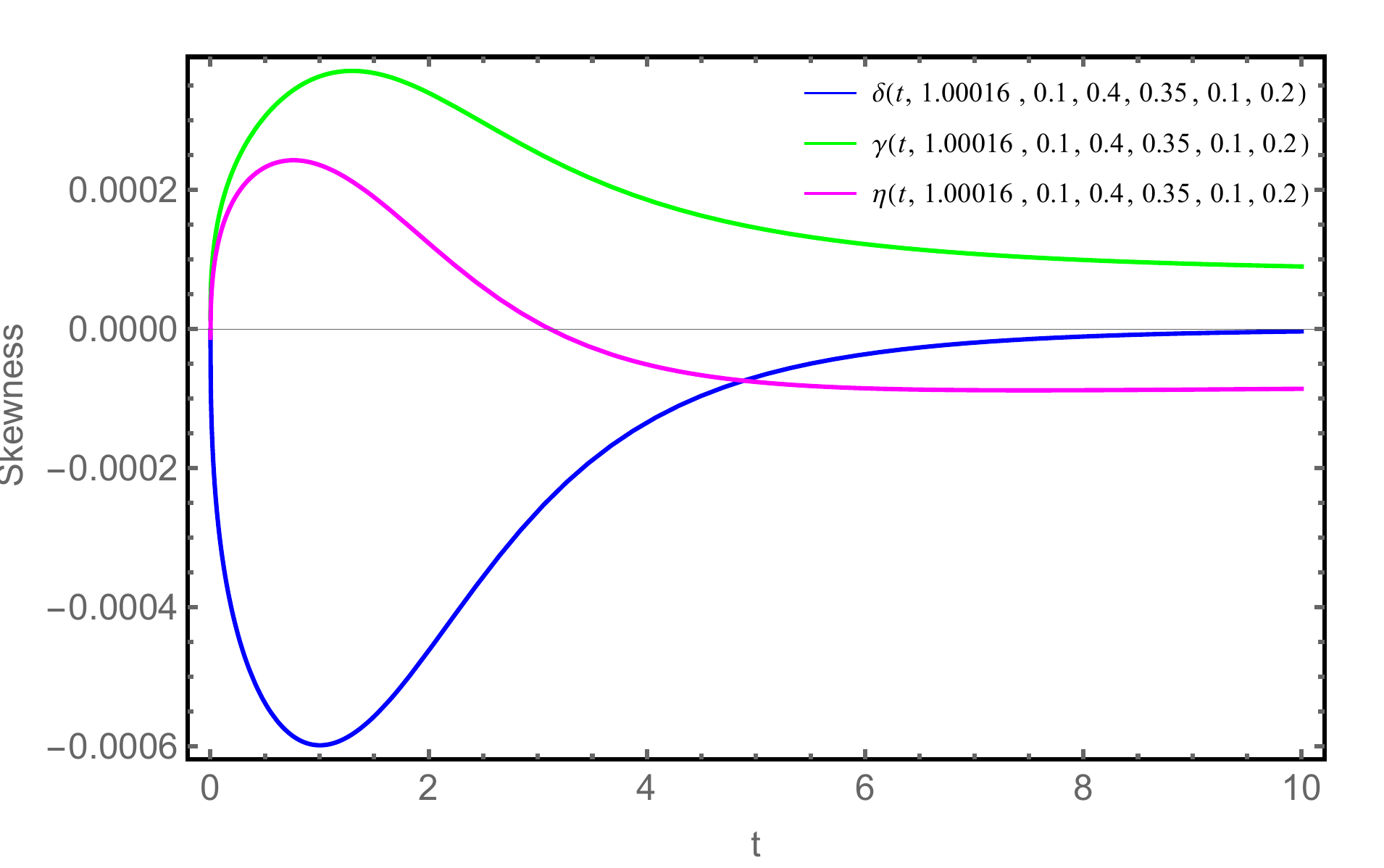} 
\caption{Skewness parameters for $k=0.1$ with $\alpha=0.35$. }
\endminipage
\minipage{0.32\textwidth}
\centering
\includegraphics[width=\textwidth]{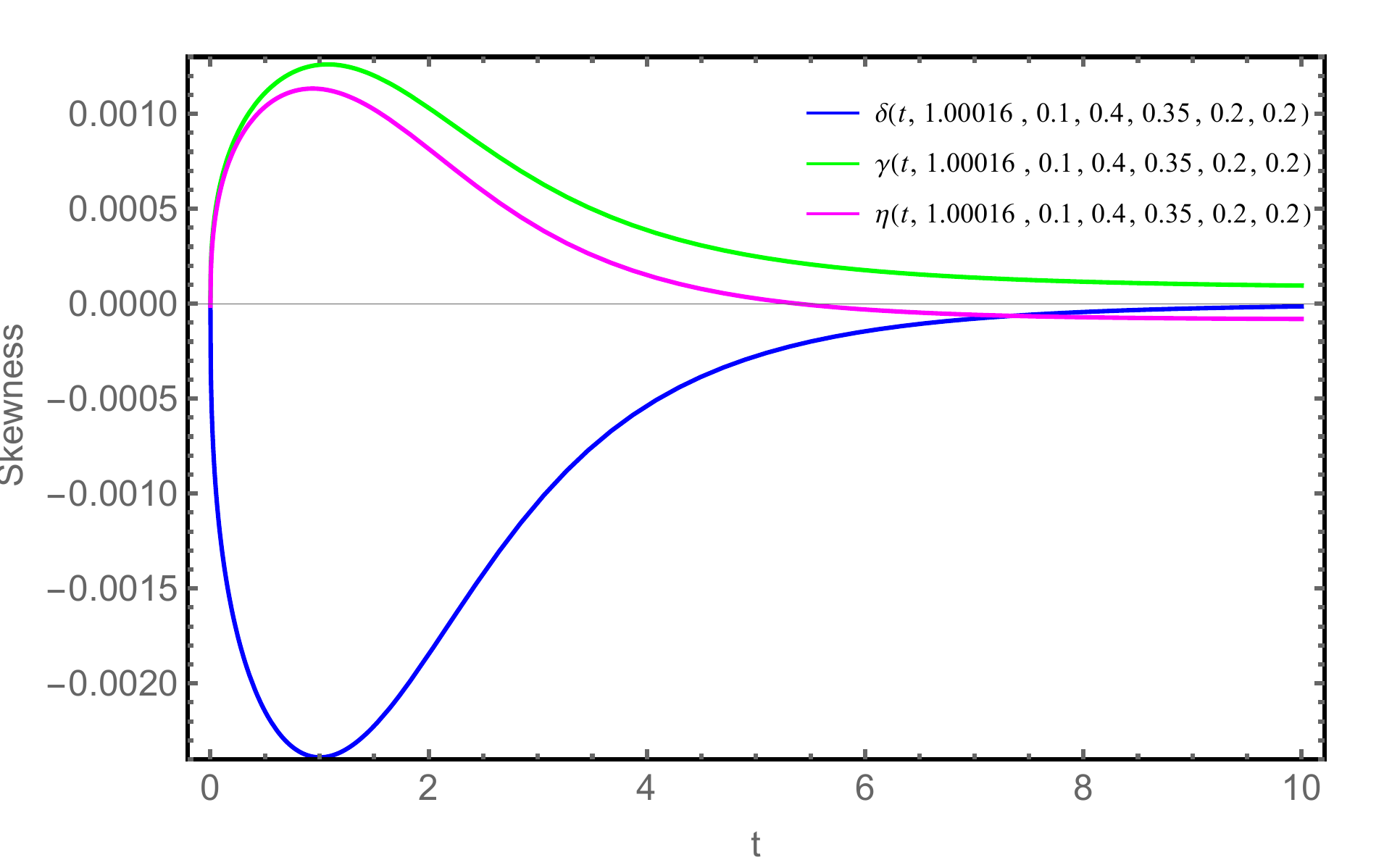}
\caption{Skewness parameters for $k=0.2$ with $\alpha=0.35$.}
\endminipage\\
\minipage{0.32\textwidth}
\centering
\includegraphics[width=\textwidth]{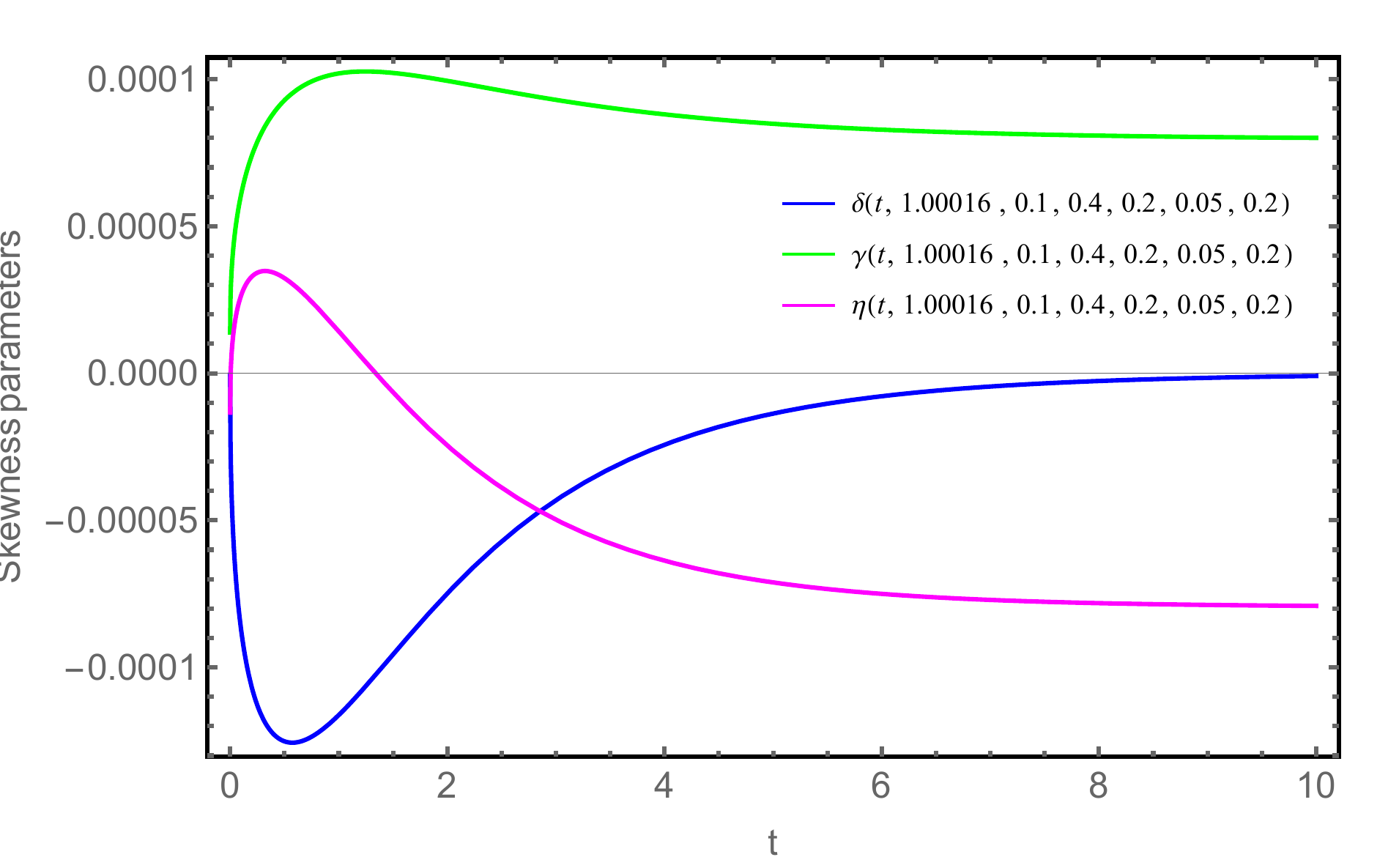}
\caption{Skewness parameters for $k=0.1$ with $\alpha=0.35$.Variation of Skewness parameters versus $t$ for $m=1.0001633,$ $ h_{1}=0.1,$ $h_{2}=0.4,$ $k=0.05$ and  $\alpha=0.2$ 
}
\endminipage
\minipage{0.32\textwidth}
\centering
\includegraphics[width=\textwidth]{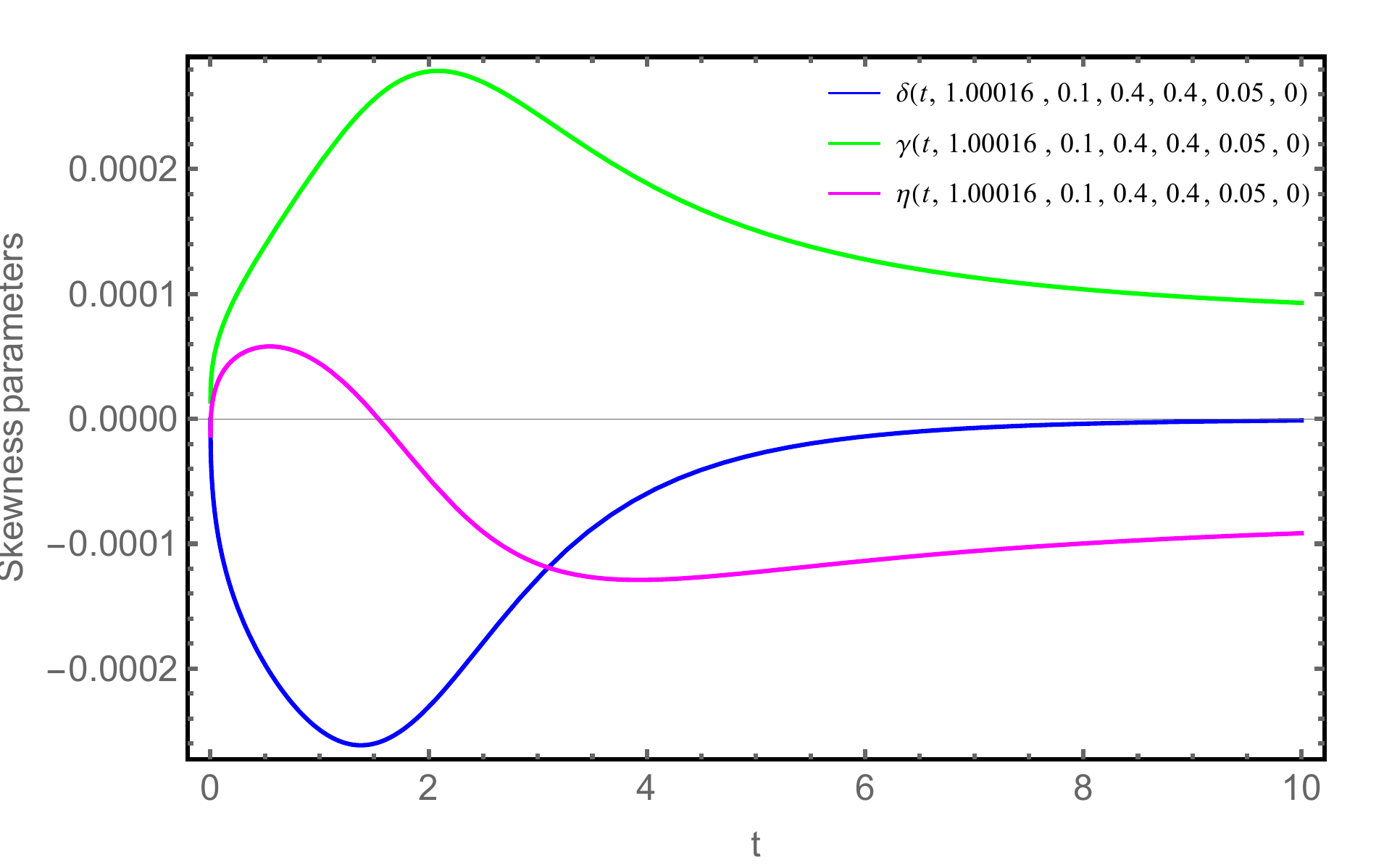} 
\caption{Variation of Skewness parameters versus $t$ for $m=1.0001633,$ $ h_{1}=0.1,$ $h_{2}=0.4,$ $k=0.05$ and  $\alpha=0.4$}
\endminipage
\minipage{0.32\textwidth}
\centering
\includegraphics[width=\textwidth]{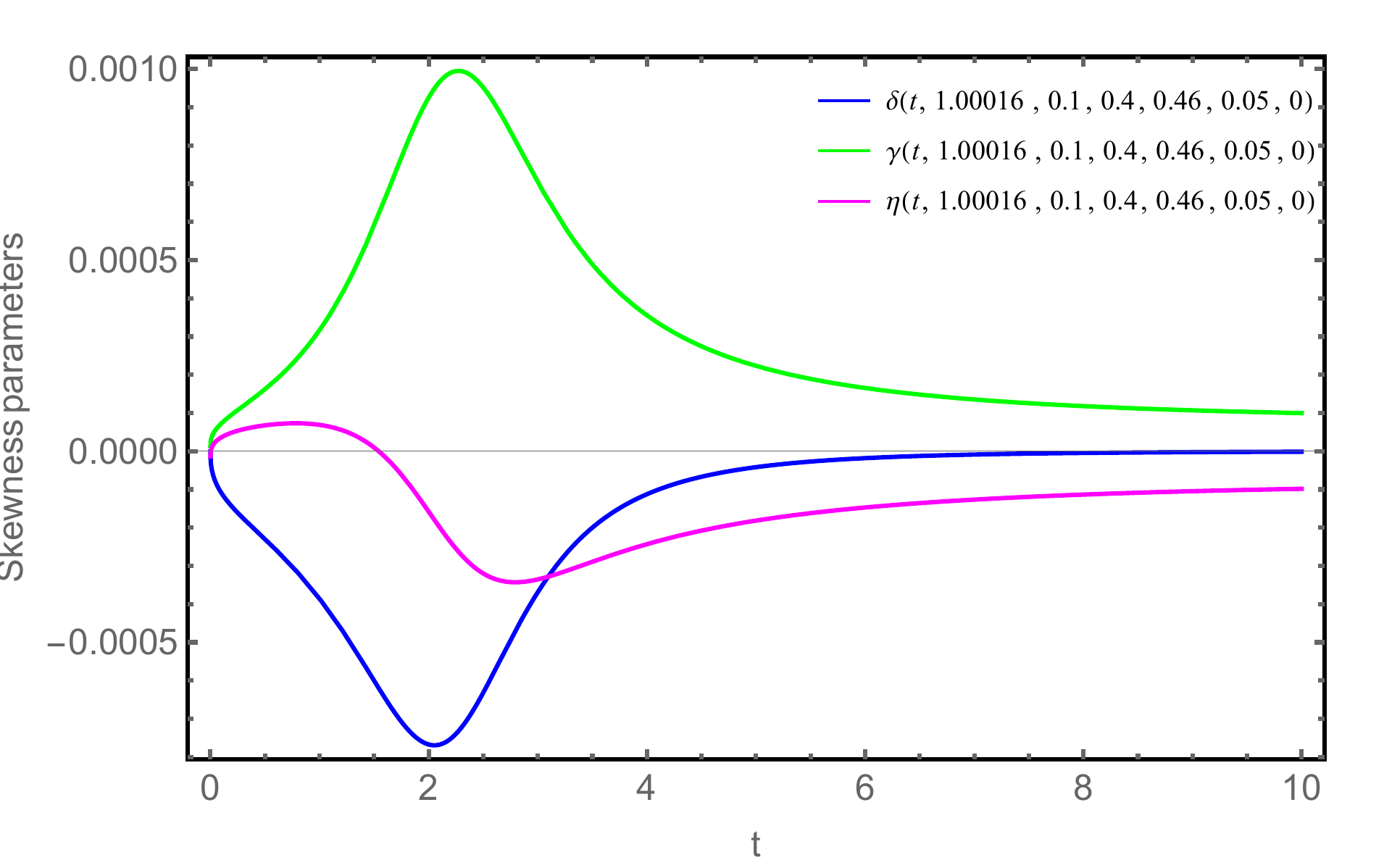}
\caption{Variation of Skewness parameters versus $t$ for $m=1.0001633,$ $ h_{1}=0.1,$ $h_{2}=0.4,$ $k=0.05$ and  $\alpha=0.46$}
\endminipage
\end{center}
\end{figure}


The evolutionary behaviour of the skewness parameters in presence of magnetic field for a constant $\alpha$ are shown in Figs.7-8.  In Figs. 9-11, the time variation of skewness parameters are shown in presence of magnetic field for different representative values of $\alpha$. When the magnetic field is switched on, the skewness parameters display interesting behaviour. The skewness parameter along the x-direction, $\delta$ which was almost non evolving in the absence of magnetic field, starts to evolve with time. It becomes negative initially and evolves to become positive at late phase. $\delta$ initially decreases steadily with time and after attaining a minimum again increases. The magnitude of $\delta$ increases with the increase in the strength of magnetic field. The skewness parameter along the z-direction, $\eta$ becomes positive in an early epoch steadily increasing initially and then decreases to become negative at late times. With an increase in the strength of magnetic field, the magnitude of $\eta$ increases. Also, the time frame (within which it vanishes) increases with an increase in magnetic field. It is interesting to note that, for a given value of $\alpha$, the $\gamma$ and $\eta$ behave alike in the presence of magnetic field. However, as is evident from Figs.9-11, $\eta$ is more sensitive to the variation of the parameter $\alpha$. 

\subsection{Existence of Big-bang and Big-rip singularity}
It can be observed that, with the representative values of the parameters, the model approaches singularity at $t = 0$. Initially, the model begins with a big bang at initial epoch and and  shows a big rip behaviour with an ending at $t = \frac{t_{0} \alpha}{h_{2}} = 31.8$.  $t_{0}$ is the present time, which is 13.82 billion years from big bang. Energy density, EoS parameter, skewness parameters diverge at $t = 0$ and $t = 31.8$. We can summarise that the model starts with a big bang at $t = 0$ with decelerating expansion, and at the late phase of expansion enters into the accelerating phase. The model isotropic at $t = \frac{t_{0} \alpha}{h_{2}}$ and indicates big rip singularity. It has finite life time. At the late phase of cosmic evolution, the universe on its expansion is in agreement with the supernovae observation and in the deceleration phase, it allows the formation of large scale structure of the universe.

\begin{figure}[h!]
\minipage{0.40\textwidth}
\centering
\includegraphics[width=65mm]{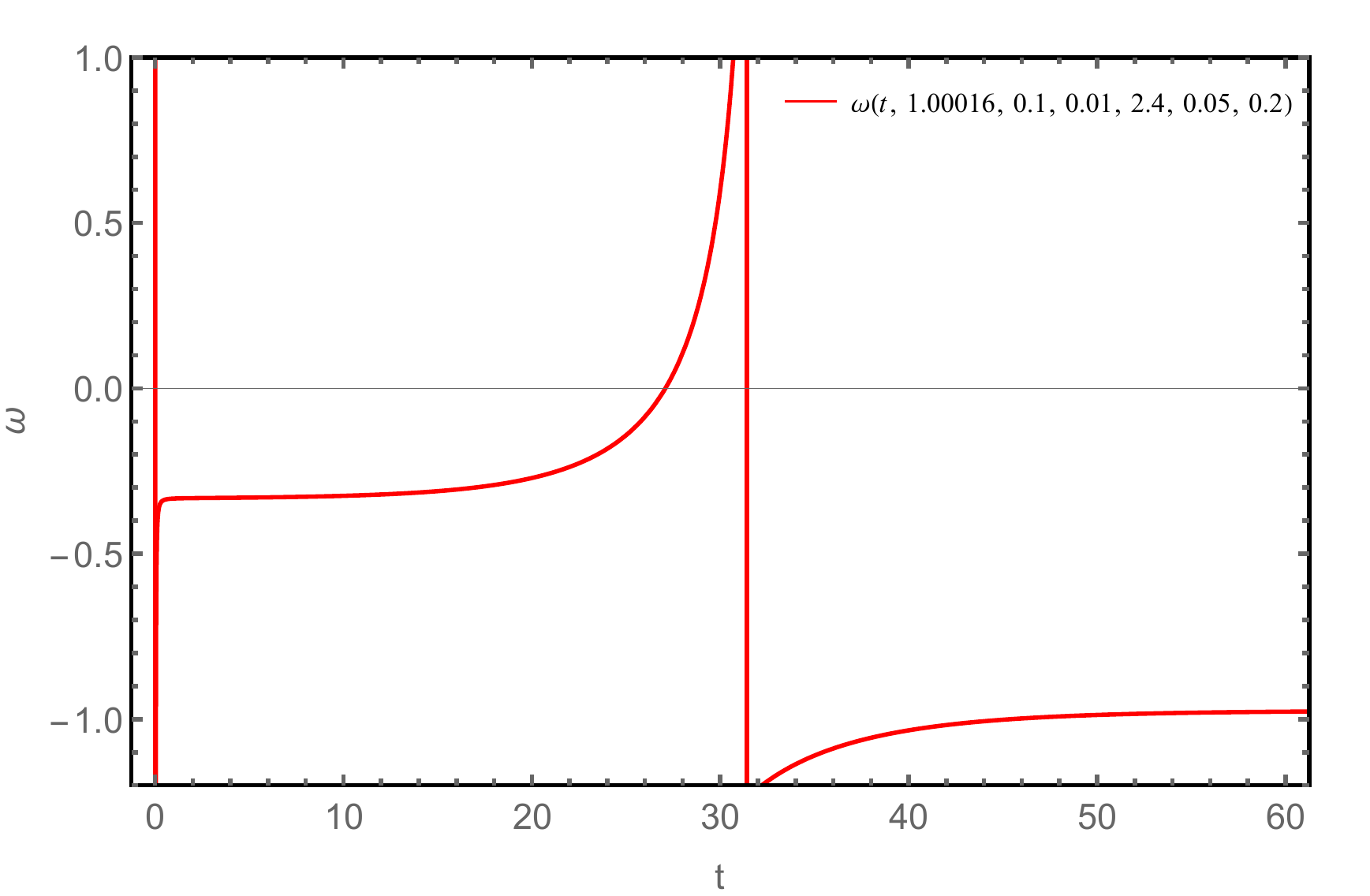}
\caption{Evolution of EoS parameter versus $t$ showing two singularities for representative values of the parameter $m=1.0001633,$ $ h_{1}=0.1,$ $h_{2}=0.01,$ $\alpha=2.4,$ $ k=0.05$ and $x=0.2$} 
\endminipage\hfill
\minipage{0.40\textwidth}
\includegraphics[width=65mm]{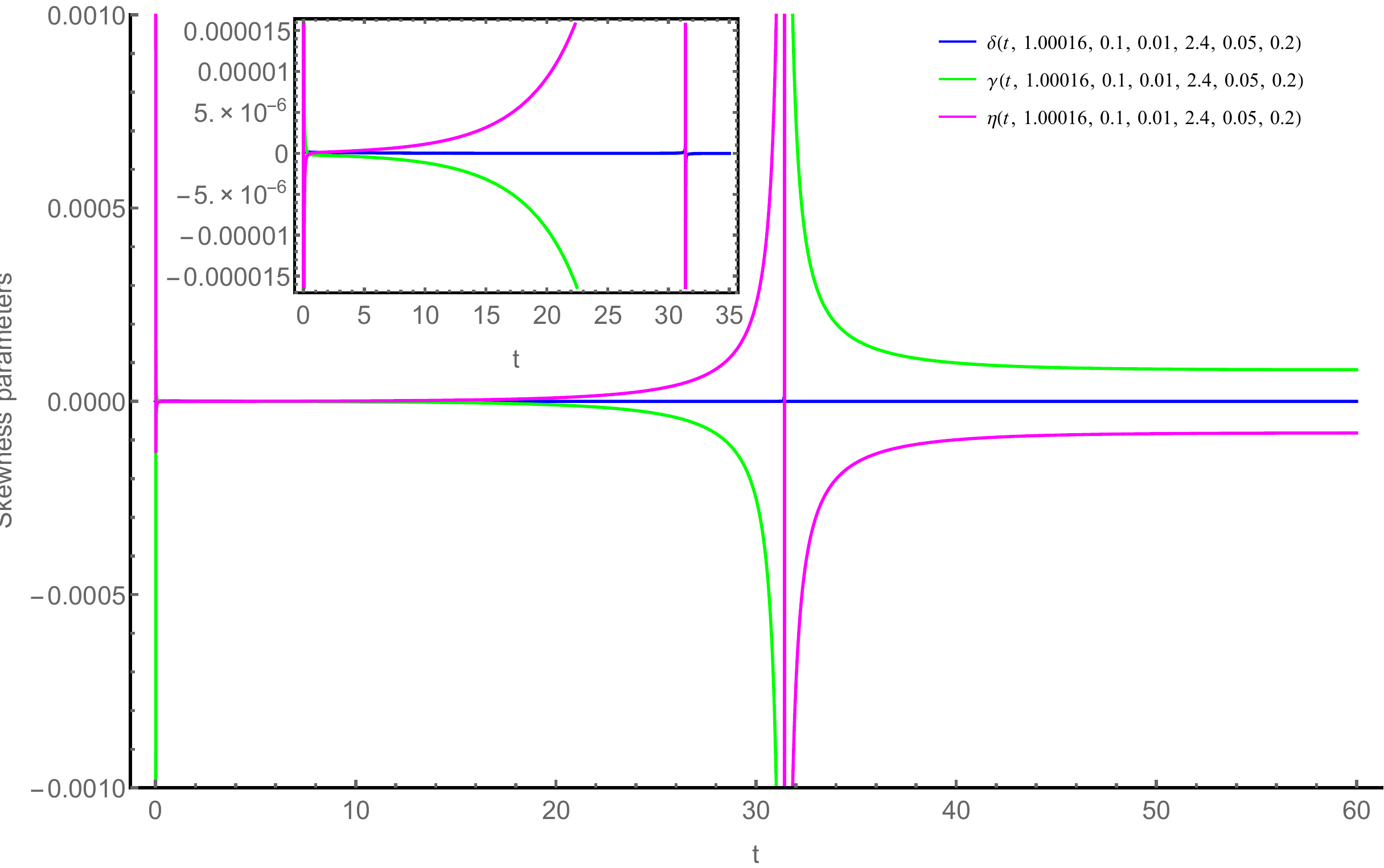}
\caption{Evolution of Skewness parameters versus $t$ showing two singularities for representative values of the parameter $m=1.0001633,$ $ h_{1}=0.1,$ $h_{2}=0.01,$ $\alpha=2.4,$ $ k=0.05$ and $x=0.2$}
\endminipage
\end{figure}

\section{Conclusion}

In the present work, in the frame work of GR, we have constructed an anisotropic DE cosmological model, with Bianchi V space time. In order to study the evolution of skewness parameters and the EoS parameters, we have developed the DE models in two fluid situation. The EoS parameter of DE evolves within the range predicted by the observations. In the present epoch, the DE dominates the universe. This may be attributed to the current accelerated expansion of the universe. The parameter $m$ is responsible for the anisotropic behaviour and the matter behaviour of the model in the sense that, if $m=1$, we get isotropic model and for $m \neq 1,$ anisotropic nature will be retained. In the present models, as expected, the matter energy density and DE density remain positive. Therefore, some energy conditions are satisfied, which in turn imply that the derived models are physically realistic. Moreover, it is shown that, there remain pressure anisotropies even at the late phase of cosmic evolution. The EoS parameter changes with cosmic time which indicates that during evolution there exists a dynamical relationship between the mean pressure of the cosmic fluid and energy density. The presence of magnetic field changes the dynamics substantially at least at the initial phase of cosmic evolution. The skewness parameters are found to be more sensitive to the magnetic field strength and the choice of model parameters.

\section*{Acknowledgement} BM and SKT acknowledge the support of IUCAA, Pune, India during an academic visit where a part of this work is done. PPR acknowledges DST, New Delhi, India for providing facilities through DST-FIST lab, Department of Mathematics, where a part of this work was done. The work of KB was supported in part by the JSPS KAKENHI Grant Number JP 25800136 and Competitive Research Funds for Fukushima University Faculty (18RI009).


\begin{thebibliography}{99} 
\section*{References}

\bibitem{Hawley98} F.J. Hawley, A. Katherine, \textit{Holcomb Foundations of Modern Cosmology}, Oxford University Press, NY (1998).

\bibitem{Ryden03} B. Ryden, \textit{Introduction to Cosmology}, Addison Wesley, CA, (2003).

\bibitem{Knop03} R. K. Knop et al., \textit{The Astrophysical Journal}, \textbf{598}, 102, (2003).

\bibitem{Riess98} A. G. Riess, \textit{The Astronomical Journal}, \textbf{116}, 1009, (1998).

\bibitem{Perlm98} S. Perlmutter et al., \textit{The Nature}, \textbf{391}, 51, (1998).

\bibitem{Riess04} A. G. Riess et al., \textit{The Astronomical Journal}, \textbf{607}, 665, (2004).

\bibitem{Ade14} P. A. R. Ade et al., \textit{Astrnomy Astrphysics}, \textbf{571}, A16 (2014).

\bibitem{Ade16} P. A. R. Ade et al., \textit{Astrnomy Astrphysics}, \textbf{594} ,A13 (2016).

\bibitem{Aghanim18} N. Aghanim et al., \textit{ arXiv:1807.06209v1}, (2018)

\bibitem{Perlm99} S. Perlmutter et al., \textit{Astrophys. J.}, \textbf{517}, 565, (1999).

\bibitem{Riess00} A. G. Riess et al., \textit{Astrophys. J.}, \textbf{536}, 62, (2000).

\bibitem{Astier06} P. Astier et al., \textit{Astron. Astrophys.}, \textbf{447}, 31, (2006).

\bibitem{Amanullah10} Amanullah et al., \textit{Astrophys. J.}, \text{716}, 712, (2010).

\bibitem{Weinberg13} D. H. Weinberg et al., \textit{Phys. Rep.}, \textbf{530}, 87, (2013).

\bibitem{Bond97} J. R. Bond et al., \textit{ Mon. Not. R. Astron. Soc.}, \textbf{291}, L33, (1997).

\bibitem{Wang06} Y.Wang, P. Mukherjee, \textit{Astrophys. J.}, \textbf{650}, 1, (2006).

\bibitem{Seljak05} U. Seljak et al., \textit{Phys. Rev. D}, \textbf{71}, 103515, (2005).

\bibitem{Adelman06} J. K. Adelman-McCarthy et al., \textit{Astrophys. J. Suppl.}, \textbf{162}, 38, (2006).

\bibitem{Jaffe06} A. H. Jaffe et al., \textit{Phys. Rev. Lett.}, \textbf{86}, 3475, (2001).

\bibitem{Spergel07} D. N. Spergel et al., \textit{Astrophys. J. Suppl.}, \textbf{170}, 377, (2007).

\bibitem{Bamba12} K. Bamba et al. \textit{Entropy}, \textbf{14}, 2351, (2012).

\bibitem{Brevik15} I. Brevik, V. V. Obukhov,A.V. Timoshkin, \textit{Astrophys Space Sci}, \textbf{355}, 399, (2015).

\bibitem{Mishra15b} B.Mishra, S.K. Tripathy, \textit{ Mod. Phys. Lett. A}, \textbf{30}, 1550175, (2015).

\bibitem{Saadeh16} D. Saadeh et al. \textit{Mon Not Roy Astro, Soc}.\textbf{462}, 1802, (2016).

\bibitem{Fayaz17} V. Fayaz, H. Hossienkhani, \textit{Eur. Phys. J. Plus}, \textbf{132}, 193, (2017).

\bibitem{Mamon17} A.Al Mamon, K. Bamba, Sudipta Das, \textit{Eur. Phys. J. C}, \textbf{77}, 29, (2017)

\bibitem{Ebrahimi17} E. Ebrahimi et al., \textit{Int. J. Mod Phys D}, \textbf{26},  1750124, (2017).

\bibitem{Bamba11} K. Bamba et al., \textit{JCAP}, \textbf{2011}, 21, (2011).

\bibitem{Bamba12a} K. Bamba et al., \textit{Astrophys Space Sci}, \textbf{342}, 155, (2012).

\bibitem{Mishra17} B.Mishra, Pratik P Ray, S.K.J. Pacif, \textit{Eur. Phys. J. Plus}, \textbf{132},429, (2017). 

\bibitem{Mishra18} B.Mishra, Pratik P Ray, S.K.J. Pacif, \textit{Advances in High Energy Phys}, \textbf{2018},6306848, (2018).

\bibitem{abbya2012} R. R. Abbyazov and S. V. Chervon, \textit{Gravit.  Cosmol.}, \textbf{18}, 262 (2012).

\bibitem{abbya2015} R. R. Abbyazov, S. V. Chervon and V. Muller, \textit{Mod. Phys. Lett. A}, \textbf{30}, 1550114 (2015).

\bibitem{Chervon} S. V. Chervon, \textit{Quantum Matter}, \textbf{2}, 71 (2013).

\bibitem{Capozziello:2010zz} V. Faraoni, S. Capozziello, \textit{Fundam. Theor. Phys.},  \textbf{170}, (2010).

\bibitem{Nojiri:2010wj} S. Nojiri, S. D. Odintsov, \textit{Phys. Rept.},  \textbf{505}, 59, (2011). 

\bibitem{Bamba:2012cp} K. Bamba, S. Capozziello, S. Nojiri, S.D. Odintsov, \textit{Astrophys. Space Sci.},  \textbf{342}, 155, (2012).

\bibitem{Bamba:2015uma} K. Bamba, S. D. Odintsov, \textit{Symmetry},  \textbf{7}, 220, (2015).

\bibitem{Cai:2015emx}  Y.F.Cai, S.Capozziello, M.De Laurentis, E.N.Saridakis, \textit{Rept.Prog. Phys.}, \textbf{79}, 106901 (2016)

\bibitem{Nojiri:2017ncd} S. Nojiri, S. D.Odintsov, V. K. Oikonomou,\textit{Phys. Rept.},  \textbf{692}, 1, (2017).



\bibitem{Akarsu10} O.Akarsu, C.B. Kilinc, \textit{Astrophys. Space Sci.}, \textbf{326}, 315, (2010).

\bibitem{Akarsu11} O. Akarsu, C.B. Kilinc, \textit{Int. J. Theor. Phys.}, \textbf{50}, 1962, (2011).

\bibitem{Yadav11} A. K. Yadav, F. Rahaman, S. Ray, \textit{Int. J. Theo. Phys.}, \textbf{50}, 871, (2011).

\bibitem {Mishra17a} B. Mishra, S.K. Tripathy, Pratik P Ray, \textit{Astrophys. Space Sci.}, \textbf{363}, 86,(2018).

\bibitem{Mishra18a} B.Mishra, Praik P Ray, R. Myrzakulov, \textit{arXiv:1801.01029v2}, \textbf{50}, 871, (2018).

\bibitem{lebla} V.G. LeBlanc, \textit{Classical and Quantum Gravity}, \textbf{14}, 2281, (1997).

\bibitem{tsaq} C.G. Tsagas, R. Maartens, \textit{ Classical and Quantum Gravity}, \textbf{17}, 2215, (2000).

\bibitem{skt2009} S. K. Tripathy, S K Nayak, S K Sahu and T. R. Routray\textit{Astrophysics and Space Science}, \textbf{321}, 247, (2009).

\bibitem{skt15} S. K. Tripathy and K L Mahanta \textit{Eur. Phys. J. Plus}, \textbf{130}, 30, (2015).

\bibitem{Kronberg:1993vk} P. P. Kronberg,\textit{Rept. Prog. Phys.},  \textbf{57}, 325, (1994).
 
\bibitem{Grasso:2000wj} D. Grasso, H. R. Rubinstein,\textit{Phys. Rept.},  \textbf{348}, 163, (2001).
 
\bibitem{Carilli:2001hj} C. L. Carilli, G. B. Taylor, \textit{Ann. Rev. Astron. Astrophys.},  \textbf{40}, 319, (2002).
 
\bibitem{Widrow:2002ud} L. M. Widrow, \textit{Rev. Mod. Phys.},  \textbf{74}, 775, (2002).
 
\bibitem{Giovannini:2003yn} M. Giovannini,\textit{Int. J. Mod. Phys.},  \textbf{13}, 391, (2004).
 
\bibitem{Giovannini:2004rj} M. Giovannini, \textit{Int. J. Mod. Phys.},  \textbf{14}, 363, (2005).
 
\bibitem{Giovannini:2006kg} M. Giovannini, \textit{Lect. Notes Phys.},  \textbf{737}, 863, (2008).
 
\bibitem{Kandus:2010nw} A. Kandus, K. E. Kunze, C. G. Tsagas, \textit{Phys. Rept.},  \textbf{505}, 1, (2011).

\bibitem{Yamazaki:2012pg} D. G. Yamazaki, T. Kajino, G. J. Mathew K. Ichiki, \textit{Phys. Rept.},  \textbf{517}, 141, (2012).
 
\bibitem{Durrer:2013pga} R. Durrer, A. Neronov, \textit{Astron. Astrophys. Rev.},  \textbf{21}, 62, (2013).
 
\bibitem{Maleknejad:2012fw} A. Maleknejad, M. M. Sheikh-Jabbari, J. Soda, \textit{Phys. Rept.},  \textbf{528}, 161, (2013).
\bibitem{Bamba:2003av} K. Bamba, J. Yokoyama,\textit{Phys. Rev. D},  \textbf{69}, 043507, (2004).
 
\bibitem{Bamba:2004cu} K. Bamba, J. Yokoyama, \textit{Phys. Rev. D},  \textbf{70}, 083508, (2004).
 
\bibitem{Bamba:2006ga} K. Bamba, M. Sasaki, \textit{J. of Cosmology and Astropart. Phys.},  \textbf{702}, 030, (2007).
 
\bibitem{Bamba:2008ja} K. Bamba, S. D. Odintsov, \textit{J. of Cosmology and Astropart. Phys.},  \textbf{0804}, 024, (2008). 
 
\bibitem{Bamba:2008xa} K. Bamba, S. Nojiri, S. D. Odintsov, \textit{Phys. Rev. D},  \textbf{77}, 123532, (2008).
 
\bibitem{Bamba:2018cup} K. Bamba, S. Nojiri, S. D. Odintsov, \textit{arXiv:1804.02275v1}, (2018).

\bibitem{Mishra18d} B.Mishra, S.K. Tripathy and S. Tarai \textit{ Mod. Phys. Lett. A}, \textbf{33}, 1850052, (2018).


 


























\end{thebibliography}
\end{document}